\documentclass{ws-ijmpcs}

\def\cm{\,\mbox{cm}}
\def\mm{\,\mbox{mm}}
\def\micron{\,\mu\mbox{m}}
\def\nm{\,\mbox{nm}}
\def\mhz{\,\mbox{MHz}}
\def\ele{\,\mbox{e}^-}
\def\pix{\,\mbox{pix}}
\def\wm2{\,\mbox{W}\,\mbox{m}^{-2}}
\def\aa{\,\mbox{\AA}}
\def\ev{\,\mbox{eV}}
\def\fs{\,\mbox{fs}}

\def\nat{Nature}

\def\pasp{Pub. Astron. Soc. Pac.}
\def\pra{Phys. Rev. {\rm A}}
\def\prd{Phys. Rev. {\rm D}}
\def\prl{Phys. Rev. Lett.}
\def\pla{Phys. Lett. {\rm A}}
\def\pnas{Proc. Nat. Acad. Sci.}
\def\sci{Science}
\def\fph{Found. Phys.}
\def\spie{Proc. SPIE}

\begin{document}

\markboth{H. Zhan}
{Which Way}

%%%%%%%%%%%%%%%%%%%%% Publisher's Area please ignore %%%%%%%%%%%%%%%
%
\catchline{}{}{}{}{}
%
%%%%%%%%%%%%%%%%%%%%%%%%%%%%%%%%%%%%%%%%%%%%%%%%%%%%%%%%%%%%%%%%%%%%

\title{Which Way?}

\author{Hu Zhan}

\address{Key Laboratory of Space Astronomy and Technology, 
National Astronomical Observatories, \\
Chinese Academy of Sciences, Beijing 100012, China\\
zhanhu@nao.cas.cn}

\maketitle

\begin{history}
\received{Day Month Year}
\revised{Day Month Year}
\published{Day Month Year}
\end{history}

\begin{abstract}
I report the result of a which-way experiment based on Young's 
double-slit experiment. It reveals which slit photons go through 
while retaining the (self) interference of all the photons 
collected. The idea is to image the slits using a lens with a 
narrow aperture and scan across the area where the interference 
fringes would be. The aperture is wide enough to separate the 
slits in the images, i.e., telling which way. The illumination 
pattern over the pupil is reconstructed from the series of slit 
intensities. The result matches the double-slit interference 
pattern well. As such, the photon's wave-like and 
particle-like behaviors are observed simultaneously in a 
straightforward and thus unambiguous way. The implication is far 
reaching. For one, it presses hard, at least philosophically, for
a consolidated wave-and-particle description of quantum objects, 
because we can no longer dismiss such a challenge on the basis 
that the two behaviors do not manifest at the same time. 
A bold proposal is to forgo the concept of particles. Then, 
Heisenberg's uncertainty principle would be purely a consequence 
of waves without being ordained upon particles.

\keywords{complementarity; double-slit experiment; particle-wave 
duality.}

\end{abstract}

\ccode{PACS numbers: 03.65.Ta; 42.25.Hz; 42.50.Xa.}

\section{Complementarity and Particle-Wave Duality}

The principle of complementarity states that complementary 
properties of a quantum object cannot be observed simultaneously.
A commonly cited example is that particles can either 
exhibit particle behavior or wave behavior in an experiment but 
not both at the same time. This is at the heart of 
particle-wave duality. Because complementarity cannot
be derived from first principle, it is regarded by many a
fundamental principle of quantum mechanics.

There are still many who wish to test complementarity. Take
Young's double-slit experiment as an example. One would try to 
determine which slit each photon goes through (i.e., which-way
information) and obtain the interference fringes at the same time. 
Along the development, it was realized that one may obtain some 
which-way information at the expense of reduced sharpness of the 
interference. Furthermore, the sum of which-way and interference 
information should not exceed the maximum available in the 
experiment,\cite{wootters1979,bartell1980,mittelstaedt1987} 
which, for instance, equals the amount of information in
absolutely accurate slit determination (thus no fringes at all) 
or complete ignorance of the slit information (thus full interference).
An inequality in terms of the fringe visibility ($\mathcal{V}$)
and the slit or path distinguishability ($\mathcal{D}$) was later
introduced:\cite{greenberger1988,jaeger1995,englert1996}
\begin{equation} \label{eq:VD}
\mathcal{V}^2 + \mathcal{D}^2 \le 1.
\end{equation}
Operationally, $\mathcal{V}$ represents the fringe contrast, 
and $\mathcal{D}$ is the normalized likelihood, in excess of a 
uniform random guess, of determining the slit correctly. 
More quantitative definitions are given in sections~\ref{sec:see} 
and \ref{sec:rei}.

One can find a number of which-way experiments in 
the literature (for a few recent examples, see 
Refs.~\refcite{kocsis2011,menzel2012}), though 
conclusive evidence of violation of complementarity has yet to 
be established. 
Chris Stubbs told me about Afshar's experiment\cite{afshar2005} 
while I was finishing this article. 
Both Afshar's experiment and mine image the double 
openings (pinholes or slits) to determine the photons' path, 
but the rest are different. In the former, a grid of thin 
wires are placed before the lens lying where the dark fringes 
of the double pinholes would be. The image of the pinholes
is only slightly affected by the wires, showing 
that the wires block and diffract only a small amount of light.
It is thus consistent with the wires being where the dark fringes 
would be. However, imagine placing a much thinner wire where 
a bright fringe would be. Much like dust on the primary 
mirror of a telescope, such a wire could easily escape 
detection in the image of the pinholes, which means that one 
cannot accurately determine the illumination pattern over the 
lens without perturbing the pinhole image significantly. 
Moreover, as pointed out in section~\ref{sec:wv}, it is not 
possible to reconstruct the illumination pattern from just one 
image of the double pinholes. Therefore, without a precise 
match of the illumination pattern with the interference 
pattern, Afshar's experiment is not a sufficient proof of 
violation of complementarity. 

Before describing my experiment, it is worth reading some 
thoughts of Bohr who first introduced the principle of 
complementarity in 1927. In the key paper published the following
year, Bohr remarked on the wave behavior of matter after 
similar comments on that of light:\cite{bohr1928}
\begin{quote}
In fact, here again we are not dealing with contradictory but with 
complementary pictures of the phenomena, which only together offer 
a natural generalisation of the classical mode of description.
\end{quote}
I imagine that Bohr was concerned about the challenge of 
particle-wave contradiction to the foundation of quantum 
mechanics. Complementarity seemed to be a necessary way out 
but not a satisfactory one by itself. 
Given that a proof from first principle was not possible, 
it was imperative to establish support of some physical
origin for complementarity. Bohr found a solution in the physics
of measurements, which helped secure complementarity and turned
the contradiction into duality. With loose ends tied up, 
he concluded the very paper with
\begin{quote}
I hope, however, that the idea of complementarity is suited to 
characterise the situation, which bears a deep-going analogy to the 
general difficulty in the formation of human ideas, inherent in the 
distinction between subject and object.
\end{quote}

What if there are no particles but only waves?

\section{Seeing the Slits} \label{sec:see}

\begin{figure}[pt]
\centerline{\includegraphics[width=10cm]{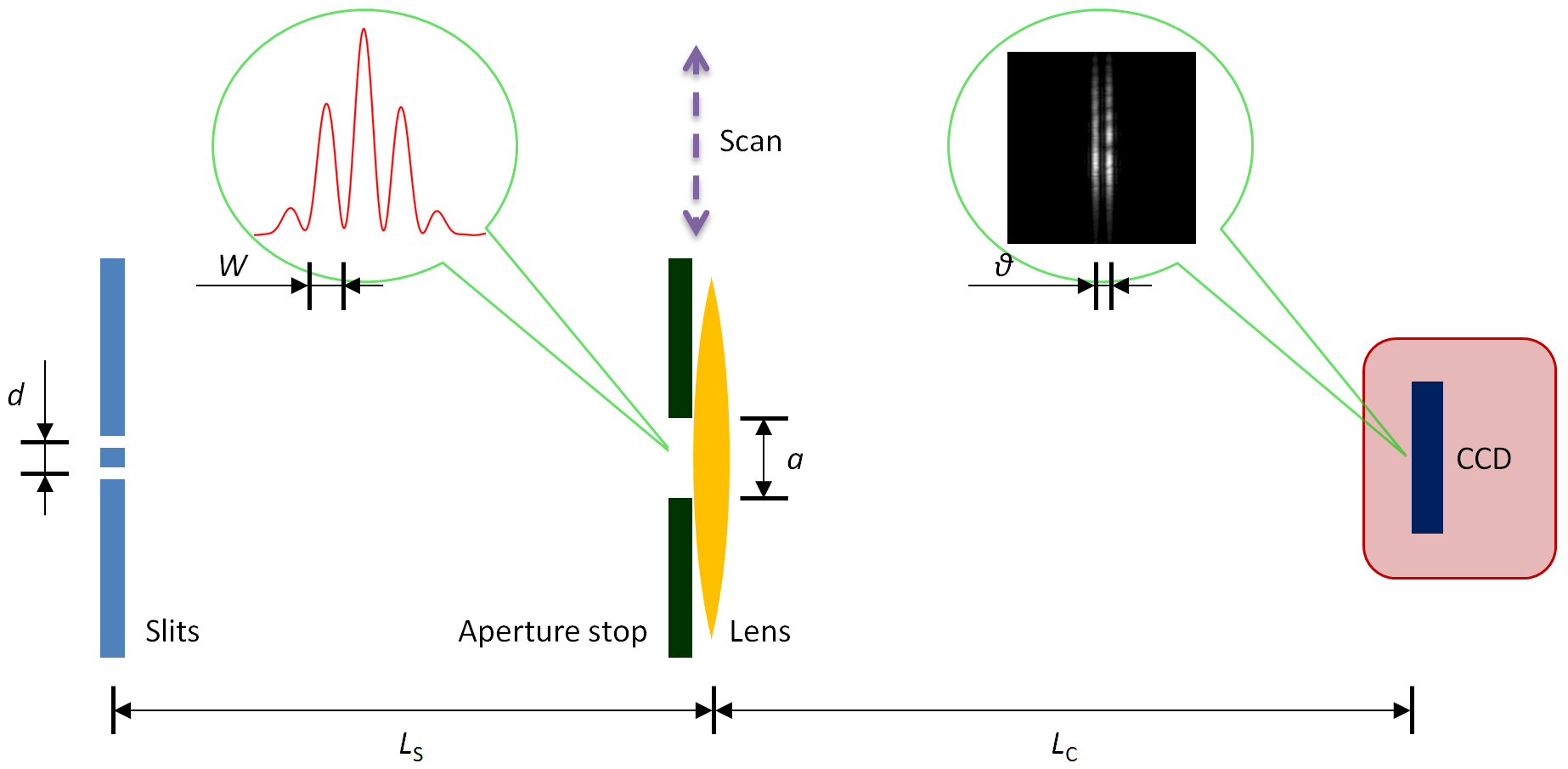}}
\caption{Experiment design. \label{fig:design}}
\end{figure}

While contemplating ways to circumvent complementarity, I imagined 
myself looking at a double-slit mask. Then, I knew how to trick the 
photons. 
Like seeing the two slits with the naked eye, the illuminated 
slits can be imaged with a camera as illustrated in 
Fig.~\ref{fig:design}. But here comes an important question: 
%\begin{quote}
\emph{would the photon stop interfering with itself as
it arrives at the lens?}
%\end{quote}

If the detector is placed right behind the thin lens, it still 
registers the interference pattern with slight optical effects.
Only when the distance between the lens and the 
detector and that between the lens and the slits roughly satisfy 
the lens equation does one get an image of two distinct slits. 
A perfect image of the slits does not reveal whether the photon 
interferes with itself or not just before hitting the lens.
If it did not, however, then some long-range interaction would be needed
to inform the photon of the details of the apparatus ahead (e.g., lens 
or screen, curvature of the lens, position of the detector, etc.) 
so that it can react accordingly before entering the lens. Neither 
electromagnetic interaction nor gravitational interaction could 
accomplish this. Introducing a new long-range force can potentially 
facilitate information delivery but cannot evade the memory
problem discussed in section~\ref{sec:ptw}. Therefore, 
my answer to the question above is ``no.''

There is still a technical problem of well resolving the slits on 
the detector and the fringes at the lens simultaneously. The 
hypothesis here is that the illumination pattern over the lens 
(more appropriately, the pupil) should be the same as the 
double-slit interference pattern, so in order to properly 
reconstruct the former the experiment should be capable of 
resolving the fringes at the position of the lens.
The characteristic scale $W$ of the fringes is given by
\begin{equation}
W = \frac{\lambda}{d}L_\mathrm{S},
\end{equation}
where $\lambda$ is the photon's wavelength, $d$ is the separation 
between the two slits, and $L_\mathrm{S}$ is the distance between the 
slits and the lens. The two slits subtend an angle 
$\theta \simeq d/L_\mathrm{S}$ to the lens. The angular 
resolution of the imaging part is roughly $\lambda/a$ with $a$ being 
the aperture width. Now there are two conflicting requirements. On  
one hand, the aperture width should be much smaller than $W$ to 
resolve the fringes well, i.e.,
\begin{equation}
a \ll W \ \ \mbox{or}\ \ a \ll \frac{\lambda}{d}L_\mathrm{S}.
\end{equation}
On the other hand, separating the two slits in the images demands 
the opposite (barely resolving the two slits is not sufficient to 
separate them to satisfaction), i.e., 
\begin{equation}
\frac{\lambda}{a} \ll \theta \ \ \mbox{or} \ \ 
a \gg \frac{\lambda}{d}L_\mathrm{S}.
\end{equation}

\begin{figure}[pt]
\centerline{\includegraphics[width=12cm]{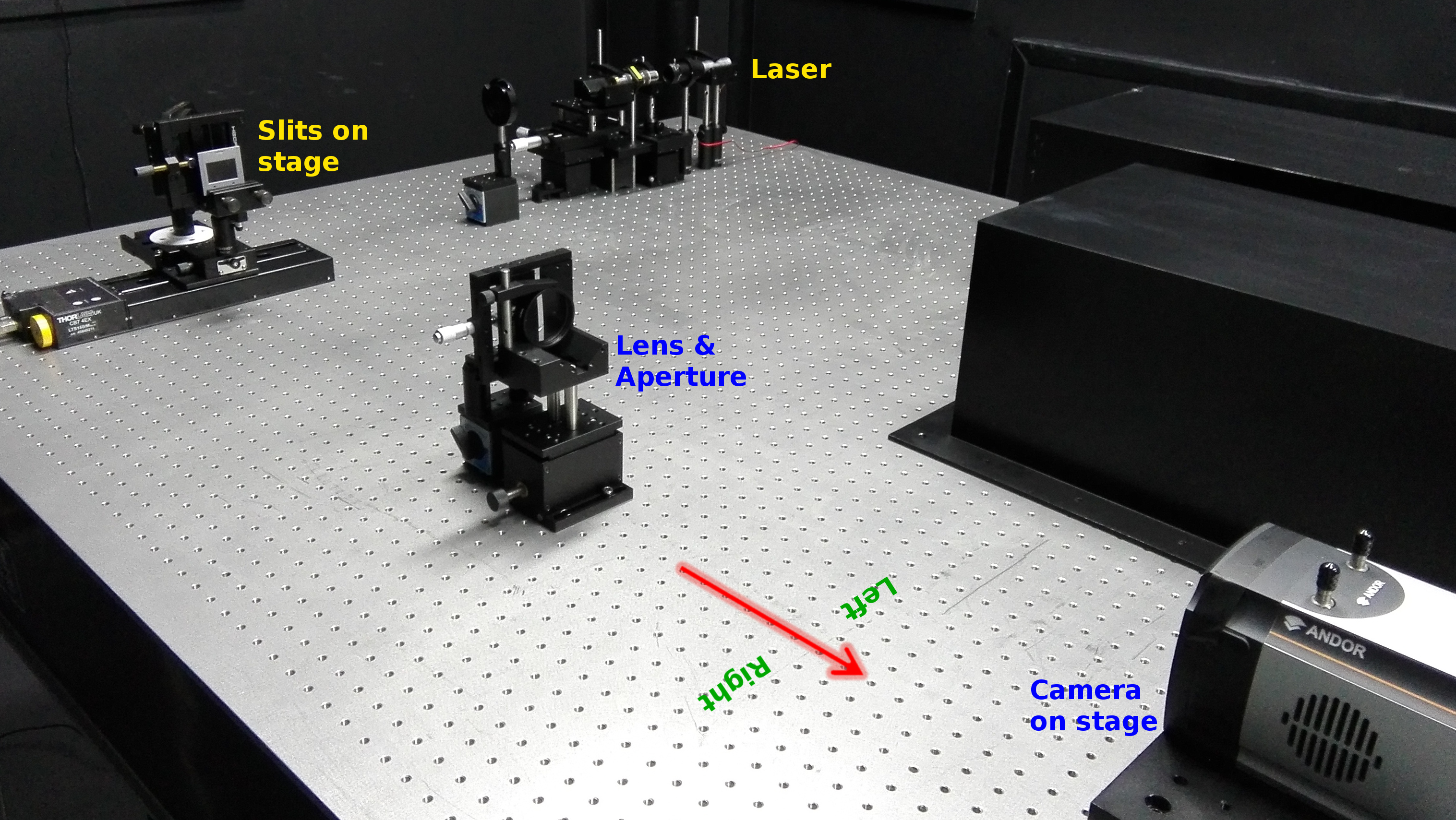}}
\caption{Experiment setup. \label{fig:photo}}
\end{figure}

For a moment, I thought quantum mechanics guarded its secret 
well. Then the technique of drizzling\cite{drizzle2002}
came to my mind, which enhances the image resolution by properly 
combining several undersampled images taken
with sub-pixel dithering. Drizzling was developed for Hubble Space 
Telescope and has become a standard practice in astronomy. It works 
because a pixel has fairly sharp boundaries capable of sampling 
spatial frequencies higher than the pixel Nyquist 
frequency\footnote{It is known as the aliasing effect in signal 
processing and is usually undesirable.}.
Hence, it is feasible to properly reconstruct the illumination 
pattern from a series of scans over the pupil plane in fine steps,
even if the aperture is wider than the fringes. One might
argue that since the slit images are taken at different times, it 
does not qualify for a simultaneous observation of the particle and 
the wave behaviors. A conceptual solution is to use a large 
number of beam splitters, lenses, aperture stops, and cameras 
to achieve the same effect of the scan while observing the slits 
simultaneously.

\begin{figure}[pt]
\centerline{\includegraphics[width=9cm]{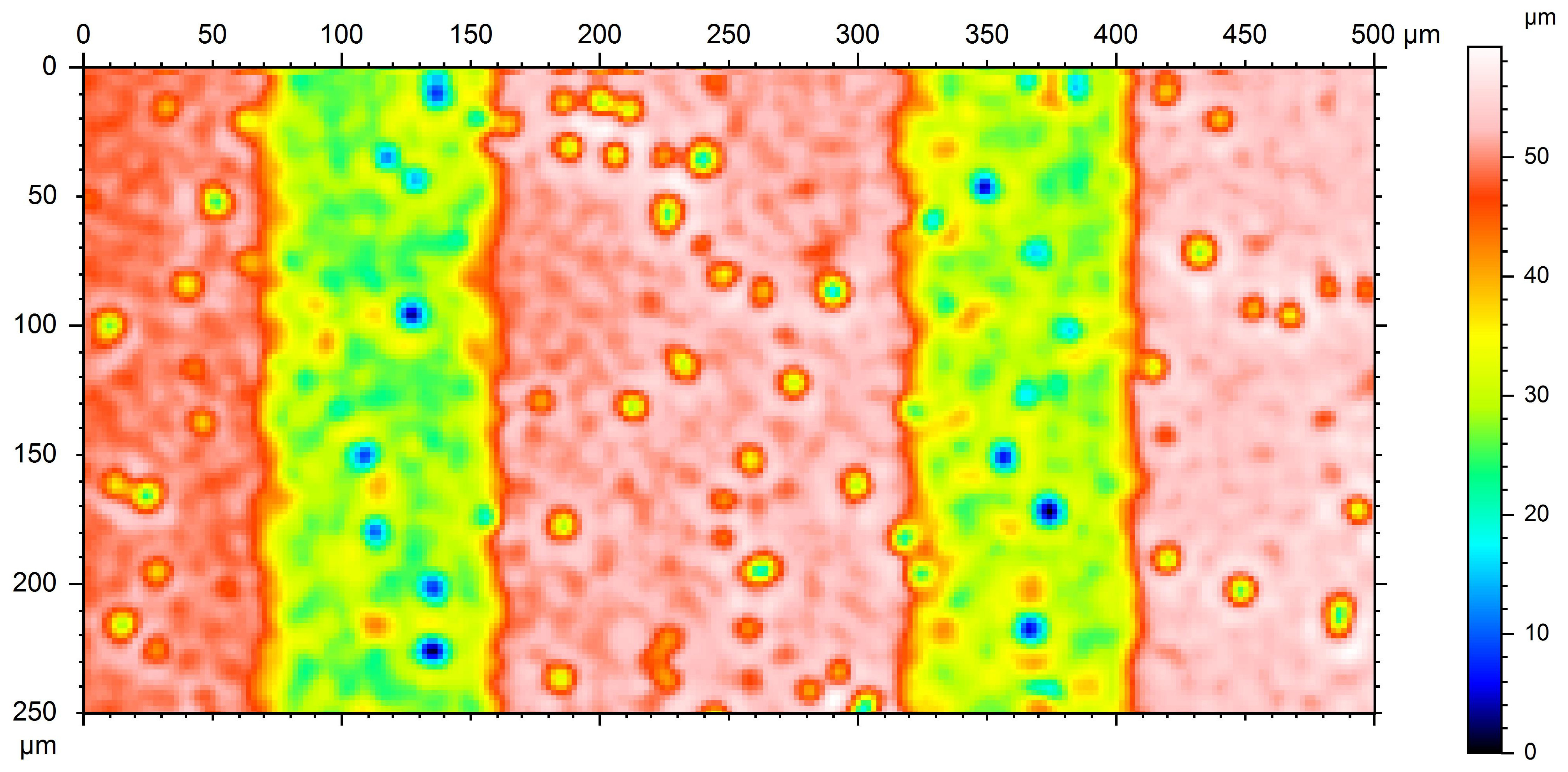}}
\caption{Surface height of a small section of the double-slit mask.
\label{fig:slits}}
\end{figure}

The experiment is prepared as shown in Fig.~\ref{fig:photo}. A red 
laser diode is used as the coherent light source. Its central 
wavelength is not precisely known, so a $650\nm$ filter (bandwidth 
$10\nm$) is added for the sole purpose of providing a definitive 
wavelength to work with. A 
spatial filter is set up to improve the beam quality. The double-slit
mask appears to be a film negative. The width ($\delta$) of each slit 
and the center-to-center distance ($d$) between the two slits are 
$89\micron$ and $248\micron$, respectively, measured from a
non-contact surface scan over a small section of the mask (see 
Fig.~\ref{fig:slits}). The aperture stop is an adjustable mechanical 
slit with a micrometer to determine its width. It is placed as close 
to the lens as possible and opens in one direction, toward the right. 
The left and right directions on the air bearing table are defined 
as one looks along the direction of propagation of the photons, so 
that they are consistent with 
those on the images taken by the camera. An example is given in 
Fig.~\ref{fig:photo}. The lens has a nominal focal length of $300\mm$.
The camera (Andor DU934P) is equipped with a $1024 \times 1024$ 
back-side illuminated deep depletion CCD
whose pixel size is $13\micron \times 13\micron$. 
At the readout rate of $1\mhz$, the readout noise is roughly $6\ele$
per pixel.

Perturbations to the lens and the aperture stop are likely to have a 
larger effect than those to the double-slit mask. Hence, instead of 
moving the former in the conceptual design of Fig.~\ref{fig:design}, 
I mount the double slits on a motorized stage (Thorlabs LTS150) with a 
minimum repeatable incremental movement of $4\micron$ and a calibrated 
absolute on-axis accuracy of $\pm5\micron$. The distance between the 
slits and the lens is $L_\mathrm{S} \simeq 58\cm$, and that 
between the lens and the camera is $L_\mathrm{C}\simeq 63\cm$. 
These distances are not accurately measured. The camera is mounted on 
another motorized stage (Thorlabs LTS300), which has the same 
specifications as the first one except for a travel range twice as long. 
The two stages move in the opposite direction at a ratio of $1:1.07$, 
so that the slits stay at the center of the detector. In this way, it 
is not necessary to flat-field the camera, and the slits and their 
diffraction wings from the aperture stop remain on the detector 
through out a scan of $30\mm$ across the illumination pattern. 

\begin{figure}
\centerline{\includegraphics[width=8cm]{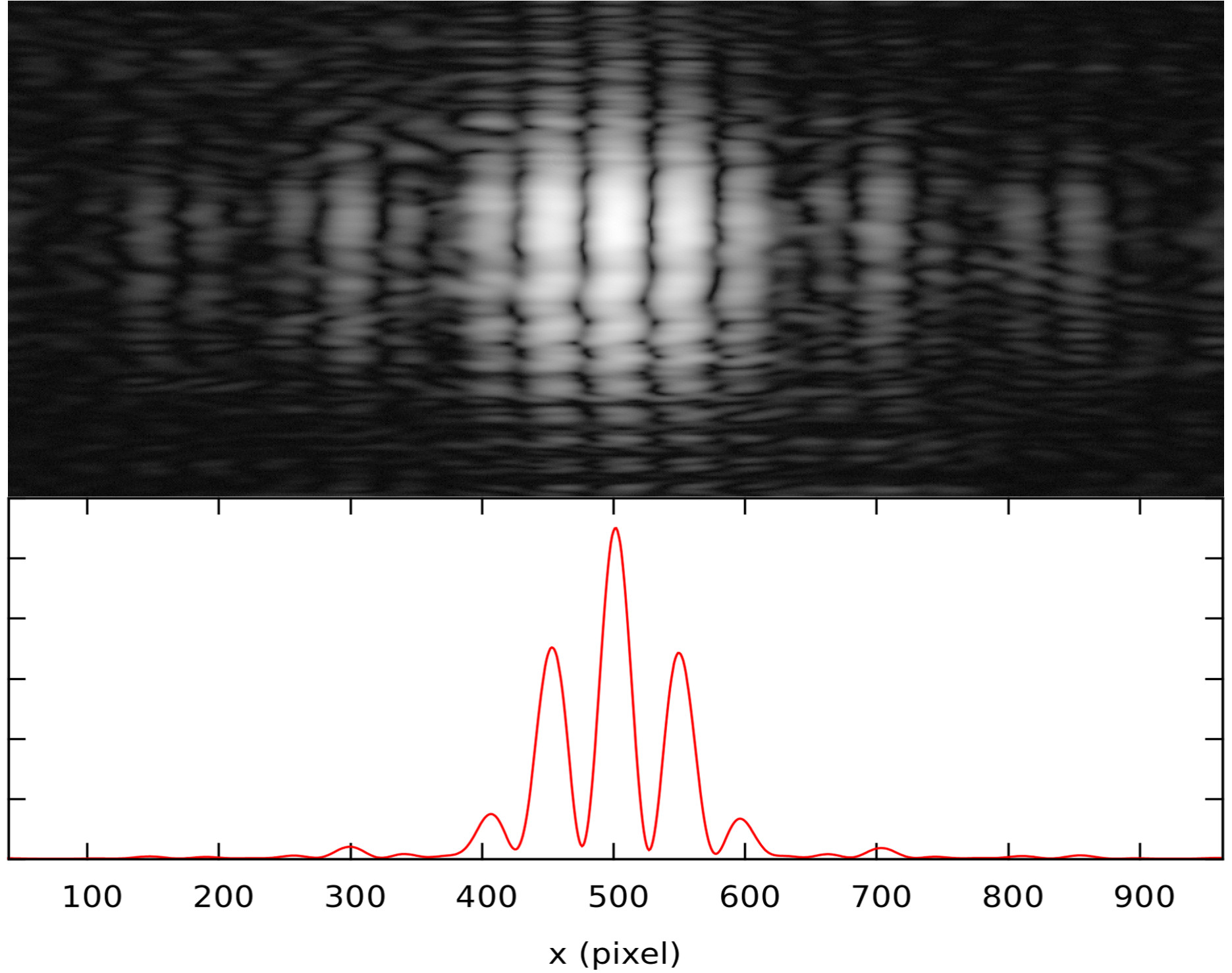}}
\caption{\emph{Upper panel}: A reference-subtracted and stacked image
of the double-slit interference pattern. The greyscale is logarithmic. 
\emph{Lower panel}: The mean flux in each column of the pixels in the 
image above.
\label{fig:fringes}}
\end{figure}

For comparison with the reconstructed result in 
section~\ref{sec:rei}, the double-slit interference pattern is 
imaged at approximately $D \simeq 25\cm$ from the slits. 
Fig.~\ref{fig:fringes} displays an average of 110 frames of the 
fringes with the mean of 105
reference images subtracted. The references are obtained in the same 
way as the fringe images are in all aspects except that the laser is 
switched off. The subtraction removes the bias, dark current, and 
darkroom background at the same time. The process of reference 
subtraction is applied to all the images in this work, and it is no 
longer mentioned hereafter. The fringes do not look as good as one 
would like because of the low quality of the slits (evident in 
Fig.~\ref{fig:slits}) and that of the laser beam. Nevertheless, the 
column-averaged fringe profile is satisfactory.

\begin{figure}
\centerline{\includegraphics[width=12.5cm]{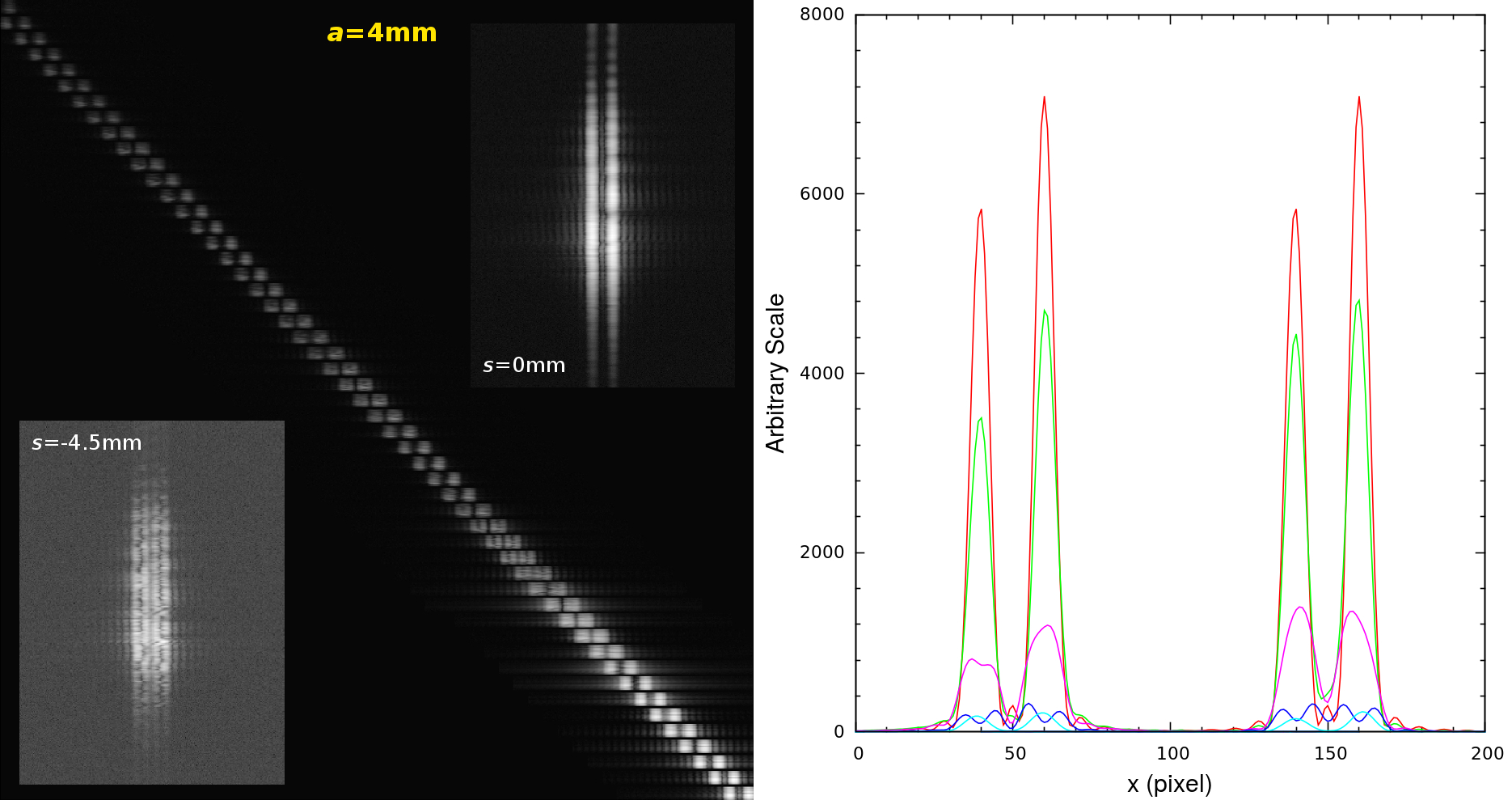}}
\caption{\emph{Left panel}: Vertically binned and cut-out images of 
the double slits at every third step in the scan sequence with the
aperture width $a=4\mm$. The top row and the bottom row correspond to 
the slits at $s=-15$ and $0\mm$, respectively. 
The results are roughly symmetric around 
the origin, so the right half of the scan is omitted. 
The horizontal shift between consecutive rows is reduced to keep all 
the slit images within the panel. The insets show the full images 
taken at $s=0\mm$, where the slits are well resolved, and $s=-4.5\mm$, 
where the slits are barely separated. The greyscales of the insets 
and the main image are all logarithmic.
\emph{Right panel}: Column-averaged intensity profiles of the slit 
images. From top to bottom, the ones centered at $x=50\pix$ 
correspond to the images taken at $s = 0$, $1.5$, $3$, $4.5$, and 
$6\mm$; the ones centered at $x=150\pix$ correspond to 
$s=0$, $-1.5$, $-3$, $-4.5$, and $-6\mm$.
\label{fig:n4cmb}}
\end{figure}

\begin{figure}
\centerline{\includegraphics[width=12.5cm]{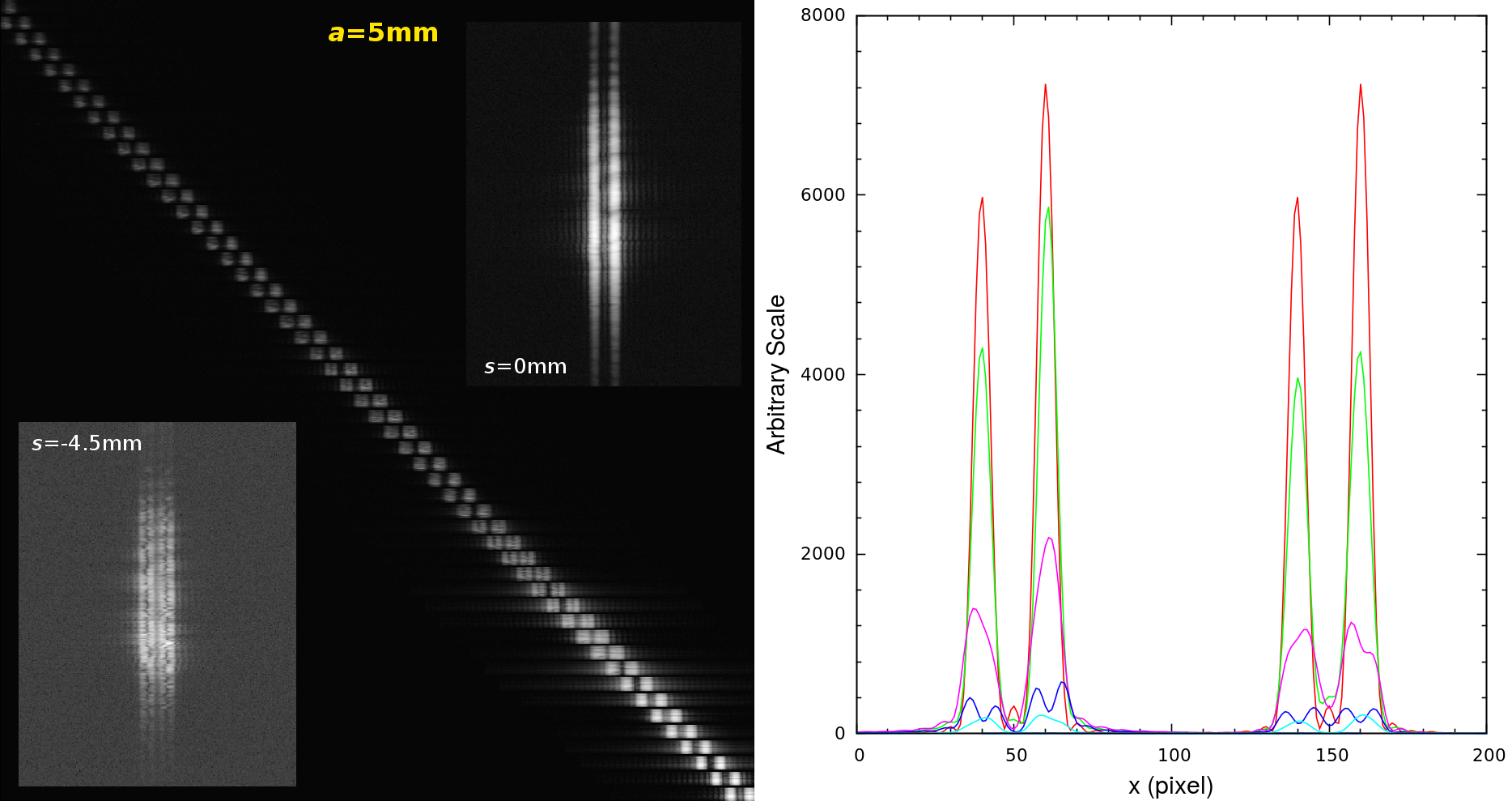}}
\caption{Same as Fig.~\ref{fig:n4cmb} but for the aperture width of
$a=5\mm$.
\label{fig:n5cmb}}
\end{figure}

Two scans are performed, one with an aperture width of $a=4\mm$ 
and the other with $a=5\mm$. The slits move from left ($s=-15\mm$) to 
right ($s=15\mm$), or, equivalently, the illumination pattern is 
scanned from right to left. The step size is $\Delta s=0.1\mm$, much 
smaller than the characteristic scale of the fringes at a distance of 
approximately $58\cm$ from the slits. Four images are taken and 
averaged at each step. The optical axis of the lens marks the nominal 
``zero'' position of the slits and that of the camera. The exposure 
time is adjusted to the aperture width so that no pixel is close to
saturation, and it remains the same throughout each scan. 
Samples of the slit images and column-averaged intensity profiles are
shown in Fig.~\ref{fig:n4cmb} ($a=4\mm$) and Fig.~\ref{fig:n5cmb} 
($a = 5\mm$). The separation between the two slits in the images 
is roughly 20 pixels or $243\micron$ in the plane of the slits, 
consistent with the slit separation measured from 
Fig.~\ref{fig:slits}.

Tagging each photon with a slit is an intrinsically 
probabilistic matter. A photon passing one of the slits can land in 
any pixel on the detector that is allowed by diffraction of the 
aperture stop, which is indeed seen in the images. As such, an
\emph{a posteriori} probability should be assigned to the photon 
going through a particular slit based 
on the location on the detector where it is registered. 
An accurate quantitative analysis is not the main concern of 
this work. It suffices to obtain a rough lower bound of the slit
distinguishability for testing the inequality in Eq.~(\ref{eq:VD}). 

Since the slits are reversed in the images, the flux to the left 
(right) of the midline between the two slits in the images should be
assigned to the right (left) slit. To avoid confusion between 
the slits and their images, I refer to the flux to the left 
(right) of the midline as the left (right) signal.
Clearly, a small fraction of 
the photons would be assigned incorrectly. Since the intensity
profile of each slit in the images is roughly symmetric around its 
center and since the distance from the midline to the center of 
either slit in each image is about 10 pixels, the amount of 
contamination to the left (right) signal would not exceed 
the total flux more than 20 pixels away to the right (left) of 
the midline. The total contamination as a fraction of the total 
signal is thus $5.1\%$ for the scan with the aperture width of 
$a=4\mm$ and $4.2\%$ for $a=5\mm$. This means that the probability 
($p$) of correct slit assignment is greater than $0.95$. The slit 
distinguishability is defined as\cite{greenberger1988}
\begin{equation}
\mathcal{D} = \left(\frac{1}{2}\right)^{-1}\left(p-\frac{1}{2}\right).
\end{equation}
Hence, this experiment achieves $\mathcal{D} \ge 0.9$.

The difference between Fig.~\ref{fig:n4cmb} and Fig.~\ref{fig:n5cmb}
may not be obvious to the eye, but the insets show narrower and more 
compact diffraction patterns with the latter as a result of the wider
aperture width in use. A further increase of the aperture width is 
likely to better separate the slits. However, I suspect that beyond 
a certain size that is determined by the double-slit interference 
pattern at the aperture stop there would be no more gain 
from increasing the aperture width, because the pupil is not uniformly
illuminated. It also means that the diffraction patterns in 
Figs.~\ref{fig:n4cmb} and \ref{fig:n5cmb} cannot always be 
described by single-slit diffraction. When they can, it is likely 
that the two edges of the aperture stop are equally illuminated.
Another idea of improvement comes from starshade\cite{cash2006}, which 
suppresses diffracted light by apodizing the sharp edge that cuts into
the light ray.

\section{Reconstructing the Interference} \label{sec:rei}

The total flux recorded by the camera at each step ($F$) is 
proportional to that passing through the aperture stop. The 
proportional constant is the system efficiency and is irrelevant 
in this work. In a discrete representation, the relation reads
\begin{equation} \label{eq:FAP}
F = A P,
\end{equation}
where $P$ is the illumination pattern over the pupil (operationally, 
it is a vector of flux values at the aperture stop in fine 
intervals), and $A$ is a matrix summing $P$ inside the aperture 
to produce the measured flux $F$ at each step (hereafter, it is 
referred to as the aperture matrix). The task is to recover
$P$ from $F$. Since $P$ cannot contain any useful 
information at spatial frequencies higher than those in $F$, the 
physical interval between two consecutive elements in $P$ should 
not be smaller than the step size of the scans, 
i.e., the reconstructed illumination pattern cannot have a 
resolution finer than that of the scans.

A problem arises immediately from the dimensions. One may 
assume that the illumination faraway from the center is too 
low to affect the reconstruction and truncate $P$ to a finite 
length. But the recorded flux vector $F$ would still have less 
elements than $P$, and the difference in physical units is the 
width of the aperture. This means that $P$ cannot be uniquely 
determined from $F$. To proceed, I truncate $P$ further to match 
the length of $F$. The associated error in reconstruction should
decrease as one increases the physical length of the scans.

Since the scans are carried out in 301 steps at intervals of 
$\Delta s = 0.1\mm$, the recorded flux $F$, the recovered 
illumination pattern $P$, and the aperture matrix $A$ have 
dimensions of 301 or $301\times 301$ 
as appropriate. The aperture widths of $4\mm$ and $5\mm$ then 
correspond to 40 and 50 elements in $A$, respectively, i.e.,
\begin{equation} \label{eq:A4}
A_{ij}(4\mm) = \left\{\begin{array}{lc}
  1 & \quad i - 20 < j \leq i + 20  \\
  0 & \quad \mbox{else}  \end{array} \right. 
\end{equation} 
and 
\begin{equation} \label{eq:A5}
A_{ij}(5\mm) = \left\{\begin{array}{lc}
  1 & \quad i - 20 < j \leq i + 30  \\
  0 & \quad \mbox{else.}  \end{array} \right. 
\end{equation} 
The same condition ``$i - 20 < j$'' appears in both 
Eq.~(\ref{eq:A4}) and Eq.~(\ref{eq:A5}) because 
the aperture stop opens toward the right. 

A straightforward way to obtain the illumination at the 
aperture stop as a function of position in the scan direction is 
to solve Eq.~(\ref{eq:FAP}). It turns out that the aperture matrix
$A(4\mm)$ is full rank only if its dimensions are multiples of 40 
or those plus one, e.g., 40, 41, 80, 81, and so on. Similarly, the 
magic numbers for $A(5\mm)$ are 50, 51, 100, 101, and so on. 
The dimensions of $301\times 301$ would not allow a unique 
determination of $P$ with $A(4\mm)$, but the degeneracy can be lifted
by incorporating $A(5\mm)$: 
\begin{equation} \label{eq:PAF}
\hat{P} = \begin{bmatrix} A(4\mm)\\ A(5\mm) \end{bmatrix}^{+}
    \begin{bmatrix} F(4\mm)\\ F(5\mm) \end{bmatrix},
\end{equation}
where the symbol ``+'' denotes pseudoinverse, and the measured
fluxes through the aperture stop have been scaled by their exposure times. 
Eq.~(\ref{eq:PAF}) is essentially a least-square estimate of 
$P$.

\begin{figure}
\centerline{\includegraphics[width=12cm]{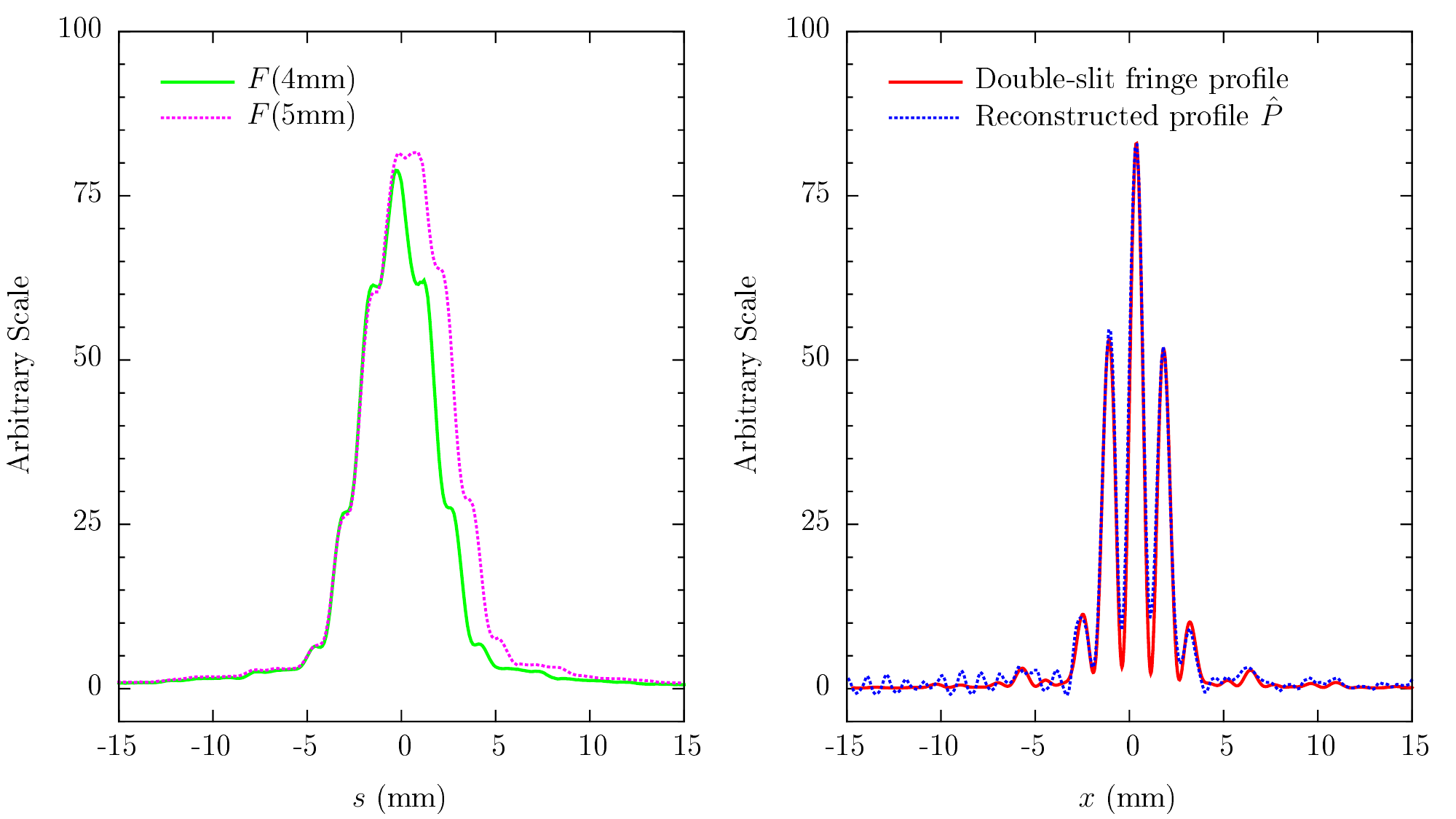}}
\caption{\emph{Left panel}: The measured fluxes as functions of the 
position in the scan sequence for the aperture widths $a = 4\mm$ 
(solid line) and $a = 5\mm$ (dotted line). 
\emph{Right panel}: The profile of the double-slit interference 
pattern from Fig.~\ref{fig:fringes} (solid line) and that of the 
reconstructed illumination pattern $\hat{P}$ (dotted line) from the 
measured fluxes in the left panel. The reconstructed profile has been 
smoothed with a Gaussian kernel that has an rms of $0.15\mm$.
\label{fig:refringes_tot}}
\end{figure}

The measured fluxes with the aperture widths $a=4\mm$ 
(solid line) and $5\mm$ (dotted line) are shown in the left panel of 
Fig.~\ref{fig:refringes_tot}. As a combined result of the $1\mm$ 
difference between the aperture widths and the fact that the 
characteristic width $W$ of the fringes at the aperture stop is a few 
times (but not much) narrower than the aperture widths, 
the flux profile $F(5\mm)$ is roughly $1\mm$ wider than $F(4\mm)$. 
The profile of the reconstructed illumination pattern $\hat{P}$ 
(dotted line) is presented in the right panel of 
Fig.~\ref{fig:refringes_tot}
along with that of the directly imaged fringe profile 
scaled from Fig.~\ref{fig:fringes} (solid line). The horizontal 
scaling factor equals $0.030\micron/\pix$, which is the pixel size 
($13\micron$) multiplied by the ratio of the distance between the slits 
and the lens ($L_\mathrm{S}\simeq 58 \cm$) to that between the slits and 
the camera ($D\simeq 25\cm$) when the interference pattern is directly
imaged. The vertical scaling is determined by matching the height of 
the central peaks of the two profiles. The only parameter that is free 
to adjust is the horizontal shift between the two curves. It is 
remarkable that the two profiles match well in the central 
region without any other tweaks. I would like to mention that even a
slight alteration to the aperture matrix, e.g., misrepresenting the 
aperture widths by just one element ($0.1\mm$ in physical units), or 
mis-aligning one aperture with respect to the other by one element,
would result in fairly noticeable deterioration to the 
reconstructed illumination pattern.

The reconstruction is not reliable in 
the outskirts, where the flux level is low. The high-frequency ripples 
that are conspicuous on the left side bear the characteristic 
period of $1\mm$, which is the difference between the two aperture 
widths. I expect that much improvement can be achieved with better 
instruments, more sampling (e.g., longer scans, more aperture widths, 
and finer steps), and better control of the laboratory environment.

For interferometry experiments, the fringe visibility is given 
by the contrast of the fringes\cite{greenberger1988}
\begin{equation} \label{eq:V}
\mathcal{V} = \frac{I_\mathrm{max} - I_\mathrm{min}}{I_\mathrm{max}
+I_\mathrm{min}},
\end{equation} 
where $I_\mathrm{max}$ and $I_\mathrm{min}$ are, respectively, the 
maximum and minimum intensities of the fringes. The double-slit
interference fringes are modulated by diffraction of individual 
slits. As the width of each slit of the double slits decreases, 
more and more fringes will become similar to the one in the 
center, and eventually they will resemble interferometric fringes.
Therefore, it is reasonable in this experiment to use the 
contrast of the reconstructed central peak and troughs
to calculate the visibility. To be conservative, I estimate 
$\mathcal{V}$ with the second and third peaks and the troughs 
between them in Fig.~\ref{fig:refringes_tot}. The result is 
$\mathcal{V} \ge 0.69$, and with $\mathcal{D} \ge 0.90$ from 
section~\ref{sec:see} one finds 
$\mathcal{D}^2 + \mathcal{V}^2 \ge 1.3$. 
The inequality in Eq.~(\ref{eq:VD}) is thus violated with a 
wide margin. It is worth mentioning that the fringe visibility is
naturally reduced by unequal illumination of the two slits as
seen in Figs.~\ref{fig:n4cmb} and \ref{fig:n5cmb}.

\begin{figure}
\centerline{\includegraphics[width=12cm]{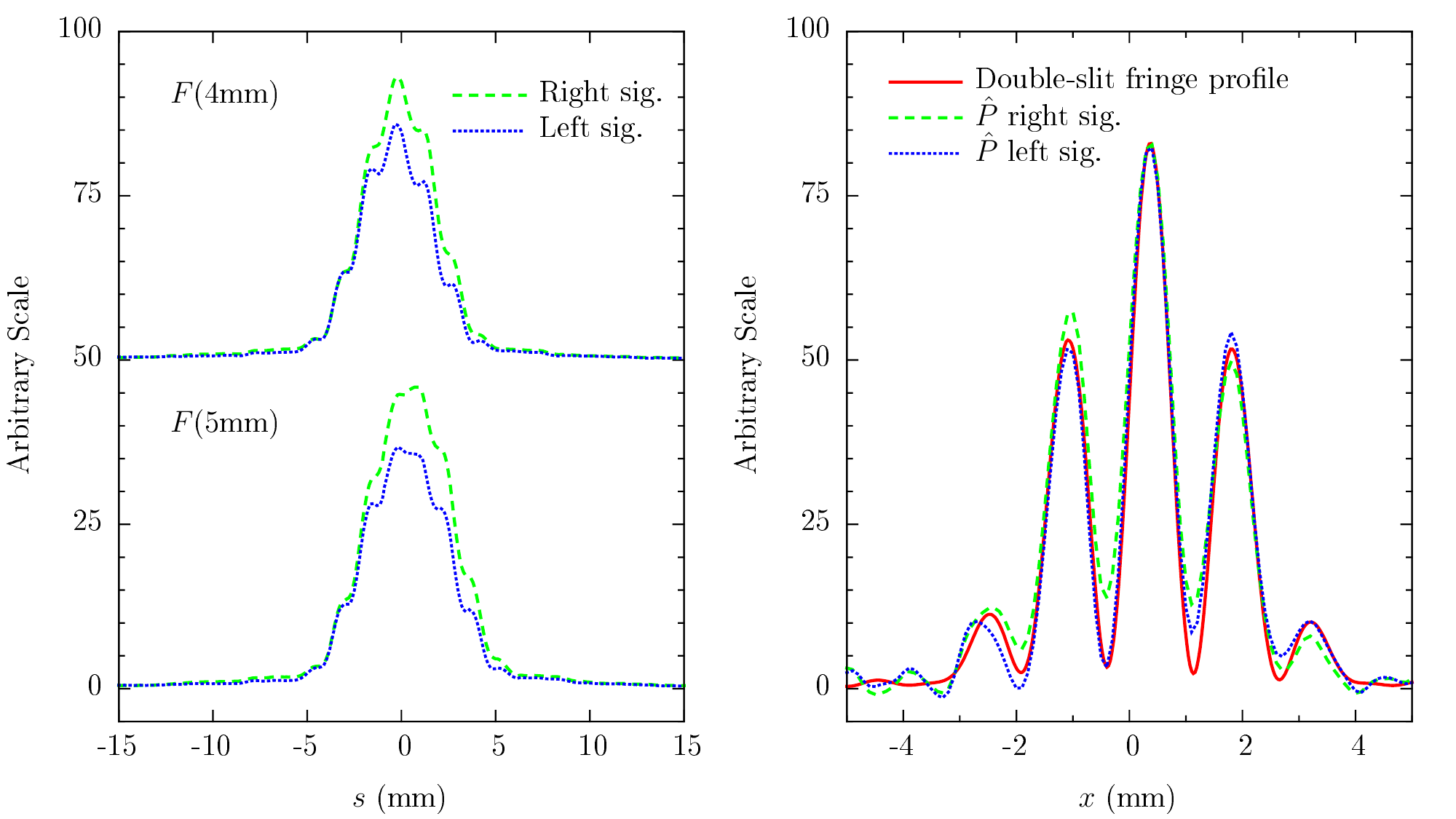}}
\caption{Same as Fig.~\ref{fig:refringes_tot} but showing results
separately for photons going through mostly the left slit (dashed 
lines) and those through mostly the right slit (dotted lines).
The fluxes $F(4\mm)$ in the left panel are shifted vertically for
clarity. \label{fig:refringes_lr}}
\end{figure}

By now one would be eager to see if there is any difference between
the reconstructed illumination pattern of the photons going through 
the left slit and that of the photons going through 
the right slit. 
The left panel of Fig.~\ref{fig:refringes_lr} displays the right 
signals (dashed lines) and the left signals (dotted lines) of the 
two scans as calculated in section~\ref{sec:see}. 
It is understood that the left (right) signal contains mostly 
photons going through the right (left) slit with slight contamination
from photons through the other slit. The right signals are higher than 
the left signals, because the two slits are not illuminated equally. 
The profile of the reconstructed illumination pattern from the right 
signal (dashed line) and that from the left signal (dotted line) 
are shown in the right panel of Fig.~\ref{fig:refringes_lr}. 
The profiles are still normalized by their central peaks. 
The most important feature is that the two profiles 
are more or less the same. However, it does not mean that a single 
slit could produce the double-slit interference pattern. 
It merely corroborates what is shown in the left panel of 
Fig.~\ref{fig:refringes_lr}: the two slits make similar
contributions to the illumination pattern everywhere across the 
pupil, albeit minor differences in shape and a more pronounced 
one in the overall intensity mentioned above.

The right panel of Fig.~\ref{fig:refringes_lr} displays an 
asymmetry between the reconstructed illumination pattern from 
the left signal and that from the right signal.
Further work is needed to determine whether it is a feature of 
physical significance or merely a 
result of an imperfect setup of the experiment.

\section{Discussion} \label{sec:dis}

My experiment demonstrates that one can determine at the same 
time which slit the photons go through with high confidence and 
recover an illumination pattern that matches the double-slit 
interference pattern. A rough estimate of the fringe visibility
and the slit distinguishability suggests that the principle of 
complementarity is violated. This result brings up a 
challenging question: \emph{what is particle-wave duality?} 
It seems that a simple solution to the conundrum is 
to forgo the concept of particles.

\subsection{Ambivalent Identities}

Although particle-wave duality is deeply entrenched in our minds, 
it seems that the particle aspect and the wave aspect are always 
discussed in different realms. There lacks a complete theory to 
explain the mechanism of particle-wave duality. It also appears 
that quantum mechanical calculations do not require mathematical 
constructs specific to the particle nature to describe particles. 
In my observation, the concept of particles is needed only 
when detection or related processes (e.g., the photoelectric 
effect) are involved. 

But, what is the defining property of a particle as opposed to a 
wave? I consider a particle to be an entity of a finite and 
more-or-less fixed size in its rest frame. A reasonable extension 
is that a particle is discrete and impermeable unless being broken 
into. Consequently, a particle cannot go through both slits in 
Young's double-slit experiment. This is the
root cause of contradiction between the particle and the wave behaviors.
From daily experience, a wave is permeable and often propagating 
into as much space as possible. These properties do not hold fast. 
Even though calculations show that the degeneracy pressure of wave 
functions can build the stiffest objects in the universe, we still 
prefer subconsciously that we are made of particles rather than waves.

A photon is always detected by a localized event, so it is natural
to consider the photon as a particle. Being a particle, the photon has 
to choose only one slit to go through. Even the thought of a photon
going through one of the slits is deeply disturbing -- how could the 
simple photon sense and react to the other slit at a macroscopic 
distance away from itself? The epistemic doctrine of ``measurement of 
the particle behavior of a quantum object impairs the measurement of
its wave behavior'' circumvents the issue, but to some it is not 
satisfactory.

\subsection{Paving the Wave} \label{sec:ptw}

The photon-counting version of Young's double-slit experiment 
could have pushed the particle-wave dilemma to the frontstage. 
However, it was not designed to reveal the trajectories of the 
photons, which is considered the ultimate test 
of their particle nature. 

The scenario of self-interference was proposed to account for the 
result of the photon-counting double-slit experiment, but 
there are difficulties. Firstly, one still has to answer the 
question of how the existence of the second slit affects the photon's 
behavior. The question can be rephrased as ``would properties 
specific to the particle nature 
be required in a theory that explains the interaction 
between the photon and the two slits?'' Secondly, what is the 
difference between a photon traveling in free space and a photon 
going through one of the two slits (or through the only slit in a 
single-slit experiment, which could also invoke self-interference)?
How does the photon know whether it should interfere with itself
or not? If it should, then where (or when) should
self-interference happen? 
It cannot occur right behind the slits, because no fringes but a
projection of the two slits can be seen on a screen there. 
Along this line, one soon 
reaches a memory problem: if the photon decides whether to
interfere with itself according to its past, then it needs a 
mechanism to store the information, which could be arbitrarily 
complex in both spatial dimensions (e.g., openings in arbitrary 
shapes and numbers) and the time dimension (e.g., arbitrary number 
of consecutive slits). It does not seem possible that a particle 
as simple as the photon can memorize an infinite history.

Waves do not suffer from the memory problem. Information can be
carried in the wave form\footnote{There is no immediate need to
identify the wave form with the photon's wave function, which, 
interestingly, is not even widely accepted as a proper concept (see, 
e.g., Refs.~\refcite{bialynicki-birula1994} and \refcite{sipe1995}).}, 
which would be a function of spatial coordinates, momentum, and 
time. The wave form would be altered by the apparatus along the 
way, analogous to the wavefront in optics. The difference between 
a photon going through double slits and another one through a 
single slit would be solely in the wave form  after they pass 
their respective slit masks. Since the photon wave could span 
across the two slits, there is no need to invent a new type
of interaction for the wave to sense both slits. It would just 
be reconditioned by the slits. 
Waves can evolve while propagating, so some distance 
after the slits would be needed for fringes to develop. 

Strictly speaking, we do not detect 
photons directly, not even with eyes; we only detect very localized 
effects of photons. Such detection ultimately involves absorption, 
which is accurately described by the interaction between two 
fields in quantum electrodynamics: 
a radiation field and a bound charged particle field without 
requiring representations specific to the particle nature 
(let us hold the question of whether electrons, 
protons, etc. are particles for the moment).
Therefore, a pixel registering a photon, no matter how small the 
pixel is, is not a sufficient proof that the photon is a particle.

One would still wonder, if the photon is really a wave that can 
spread over a large area, how can it hit just a single pixel? 
We might borrow the idea of wave function collapse in 
quantum measurement theory, but I think much
work is needed to fully understand the process of collapsing. 
My conjecture is that once the photon wave arrives at the 
detector, the best matching atom (wave) would absorb the 
photon in a runaway process, excluding the possibility of 
being absorbed by another atom elsewhere at the same time. 
The atoms in the detector would be in random phases, so they 
would sample the photon wave at random individually but still
produce the double-slit fringes collectively. 
If no atom is ready to absorb the photon, reflection or 
transmission occurs. This scenario can be interesting when 
identical targets are prepared coherently.

\subsection{The Photoelectric Hurdle}

It is widely accepted that the photoelectric effect anchors the 
particle nature of the photon. Since it is an absorption process,
the same argument in section~\ref{sec:ptw} applies. 
However, it is worth delving a bit deeper and even digressing 
a little. 

Let us start with a quote from a translation of 
Einstein's words\cite{einstein1905,haar1967}:
\begin{quote}
\ldots it is quite conceivable that a theory of light involving 
the use of continuous functions in space will lead to contradictions 
with experience, if it is applied to the phenomena of the creation 
and conversion of light. \\ 
\ldots when a light ray starting from a point is propagated, 
the energy is not continuously distributed over an ever increasing 
volume, but it consists of a finite number of energy quanta, 
localised in space, which move without being divided and which can 
be absorbed or emitted only as a whole.
\end{quote}
The above statement assumes that a 
continuous function must behave continuously in all aspects.
It need not be so. The wave function of a particle in a box is 
continuous, but its energy is discrete, proportional to 
$n^2m^{-1}$, where $n$ is a positive integer, and $m$ is the 
particle mass. If we drop the concept of particles here, then 
the mass is just another parameter of the wave (function) 
much like the wavelength of the photon. Such waves can only 
exchange discrete amounts of energy in interactions with 
each other. Since, unlike the radiation field, photons themselves
do not have excitation states, we can 
make an analogy between them and the ground-state wave ($n=1$) 
in the box with the photon energy controlled by its wavelength
rather than mass. If a photon wave is ever absorbed by another 
wave, the latter has to take all the energy of the former as a 
whole.

Contrary to common believes, classical physics would practically
prohibit the photoelectric effect if energy is continuously imparted 
on electrons in a metal. Let me use industrial CO$_2$ lasers 
as an example. They can easily reach a power density of $10^{12}\wm2$, 
far powerful than any light source obtainable in the 1900s; for 
comparison, the solar constant is a feeble $1361\wm2$. Assuming 
that the lattice size of the metal is roughly $5\aa$, one gets an 
incident laser power of $2.5\times 10^{-7}\,\mbox{W}$ on each 
lattice at the surface of the metal. This is the upper limit of 
power available to any electrons in the lattice. In the Drude 
model\cite{drude1900}, electrons are constantly colliding with 
each other (and ion cores) in the metal, so that they are in good 
thermal equilibrium. An electron must gather enough energy in a
time scale shorter (or, at least, not much longer) than the mean 
time between collision ($\tau \sim 1$-$10\,\fs$) to escape from 
the metal surface before subsequent collisions thermalize it with 
the rest of the metal. However, even if we take the rough upper 
bound of $\tau$ as the absorption time scale, the maximum 
amount of energy the electron can get in this time is only 
$2.5\times10^{-21}\,\mbox{J}$ or $0.016\ev$, still a few hundred 
times smaller than work functions of many common metals. 
Before one dials up the power of the laser 
trying to kick the electrons out of the metal, the heat flow of
the energized electrons would be already so intense that the spot 
lit by the laser would melt or vaporize -- this is how laser beam 
machining works.

The analysis above suggests that, to produce the photoelectric 
effect, one would need a mechanism to deposit several 
electronvolts of energy on the electron in less than a femtosecond 
or so. Quantized photon energy alone, or even with an 
implicit assumption of instantaneous absorption, 
is somewhat incomplete to explain the photoelectric effect, 
because physics does not prohibit absorption of multiple photons 
by the electron before it escapes from the metal surface. 
In fact, G\"{o}ppert-Mayer predicted multiple-photon
absorption in 1931\cite{goppert-mayer1931}, which was  
confirmed 30 years later\cite{kaiser1961}. 
Absorption of multiple photons in a time less than $\tau$ could 
ruin the linear relationship between the electron's maximum 
kinetic energy and the frequency of the incident light, though 
there is little chance for such events to happen.
For example, the aforementioned CO$_2$ 
laser delivers only $0.13$ photon ($\lambda=10.6\micron$) per 
second on each lattice, and this is already several orders of 
magnitude more powerful than needed to cut through most materials.

\begin{figure}
\centerline{\includegraphics[width=7.2cm]{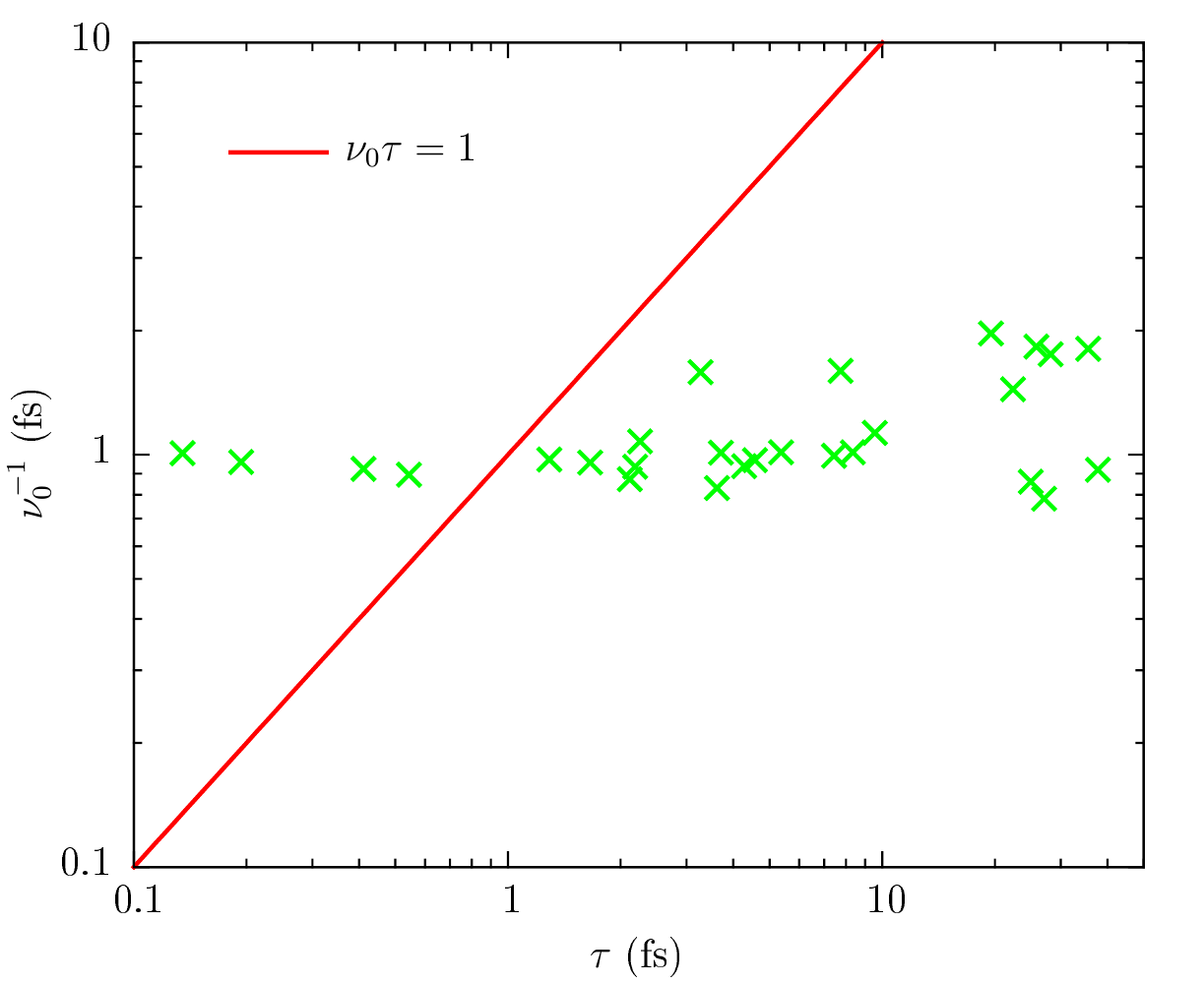}}
\caption{Comparison of the reciprocal of the minimum photon 
frequency to produce the photoelectric effect and the mean time 
between collisions for various metals. From left to right, the 
symbols correspond to Mn, Bi, Hg, Sb, Pb, Ga, Fe, Sn, Tl, Ba, Be, 
In, Nb, Zn, Cd, Al, Sr, Li, Mg, Cs, Ca, Cu, Rb, Au, Na, K, and Ag.
The region well above the line of $\nu_0\tau=1$ is disfavored by
the photoelectric effect.
\label{fig:tau_nu}}
\end{figure}

In summary, the concept of energy quanta is compatible with 
continuous waves.
It is important to recognize the two time scales in the
photoelectric effect: the mean time between collisions for the
electrons ($\tau$) and the time to absorb a photon ($t_\mathrm{ab}$, 
which may be identified with the time for the wave function to 
collapse). The latter does not have to be strictly shorter than the 
former for the photoelectric effect to take place, because the 
actual intervals between two collisions fluctuate around $\tau$. 
However, a severe penalty of the photoelectron yield would be paid 
if $t_\mathrm{ab}$ is considerably longer than $\tau$.
Thus, the photoelectric effect sets a loose 
upper bound for $t_\mathrm{ab}$. It is also reasonable to assume 
that the time for absorption should be longer than the reciprocal of 
the photon's frequency ($\nu$). Therefore, we have 
\begin{equation} \label{eq:tau_nu0}
\tau \gtrsim t_\mathrm{ab} \gtrsim \nu^{-1} \quad \mbox{and} \quad
\tau \gtrsim \nu_0^{-1},
\end{equation}
where $\nu_0$ is the minimum photon frequency to overcome the 
work function. For most metals, the work function is $2$-$6\ev$, 
and $\tau$ is $1$-$10\fs$. Hence, an order-of-magnitude
estimate for the absorption time is $t_\mathrm{ab} \sim 1\fs$. 
Fig.~\ref{fig:tau_nu} lists $\nu_0^{-1}$ against $\tau$ for 
various metals. The mean time between collisions is estimated 
from the electrical conductivity and ``free'' electron density
in the metal according to the Drude model. It is interesting 
that four metals do not conform to the last inequality in 
Eq.~(\ref{eq:tau_nu0}). These metals also 
have the lowest conductivity among the ones in the figure,
suggesting systematic inadequacies of the classical Drude model 
to describe them.

\subsection{A Wavy View} \label{sec:wv}

In retrospect, the which-way question becomes irrelevant 
if the photon is a pure wave. 

Let us examine the experiment in this work with a
wave interpretation. The photon wave is described by 
its shape\footnote{Shape here means the aspect of the wave 
form that changes with spatial coordinates, and its conjugate 
quantity is taken as momentum (but not the real momentum).}, 
momentum, and time evolution. When there is no lens 
between the camera and the slits, one gets an image of
randomly sampled wave shape. With sufficient sampling, the 
fringes appear, but the momentum information, which would
tell the path of the photon in the particle interpretation, is lost.
An ideal lens without the aperture stop would Fourier transform 
the photon wave to provide a momentum representation in the image
plane. The shape information is still encoded in the Fourier phases,
which, unfortunately, is discarded by the camera. 
As such, one cannot manipulate a single image of the two slits
in any way to recover the illumination pattern at the pupil. 
The effect of the aperture stop is a convolution of its Fourier
modes with the illumination pattern's Fourier modes. The scan shifts 
the relative phases between the two parties in the convolution.
After taking the data of many shifts, one recovers shape 
information to satisfaction. Fig.~\ref{fig:refringes_lr} is quite
interesting. It shows that the two momentum components of the 
photon wave have very similar (perhaps the same) shapes, 
analogous to self-interference occurring regardless which slit a
particle photon goes through. 

How about other particles? The same arguments should apply. 
If one agrees that neither localized events nor quantized energies
constitute a necessary condition for declaring detection of 
particles rather than waves, then there is little evidence or 
need for ``particles'' to be particles. 
I therefore propose to describe quantum objects in a wave-only 
representation. Consequently, the concepts of interference and 
diffraction are  not needed anymore, because they can be 
described by waves passing openings of different sorts.
Another benefit is that we do not have to impose Heisenberg's 
uncertainty principle on particles if they do not exist. 
One can prove the uncertainty principle mathematically for
wave functions, but not particles with bare particle properties.

So, are we waves after all?

\section*{Acknowledgments}
I would like to thank Charling Tao and Chris Stubbs for useful
discussions.

\section*{Epilogue}
Although the particle picture of light has difficulty in explaining
Young's double-slit experiment, it is so well shielded by 
complementarity that an explanation has long been deemed unnecessary. 
After all, the wave picture of light is not without its own problem. 
This work tries to lift the shield and makes an attempt to 
reconcile the wave picture and energy quanta heuristically, though
I think quantum electrodynamics has already provided a formal 
solution. It is certainly premature 
to discard the concept of particles at this point, but the need for 
a full understanding of particle-wave duality is nonetheless 
outstanding.

%\bibliographystyle{ws-ijmpcs}
%\bibliography{ref}

\end{document}